# Effect of hydraulic conductivity and permeability on drug distribution, an investigation based on a part of a real tissue

Masod Sadipour[1], Mohammad Masoud Momeni[2], Majid Soltani[2,3,4,5,*]

[1] Department of Mechanical and Materials Engineering, University of Denver, Denver, Co, USA

[2] Department of Mechanical Engineering, K. N. Toosi University of Technology, Tehran, Iran

[3] Department of Electrical and Computer Engineering, University of Waterloo, Waterloo, ON, Canada

[4] Waterloo Institute for Sustainable Energy (WISE), University of Waterloo, Waterloo, ON, Canada

[5] Advanced Energy Initiative Center, Multidisciplinary International Complex, K. N. Toosi University of Technology, Tehran, Iran

* Corresponding Author. Email: msoltani@uwaterloo.ca

**Abstract:**

In this study, a computational simulation is employed to place two essential parameters, the permeability of vessels and hydraulic conductivity, under assessment. These parameters impact the movement of drug particles through an vessels, normal and tumoral tissue to examine the concentration of nanoparticles, interstitial pressure and velocity. To provide a geometric model detailing the capillary network under normal and tumoral tissue's conditions, the geometry is extracted via real image processing. Subsequently, the real conditions were considered to solve the equations pertaining to drug transport and intravascular and interstitial flows in the tissue. The results showed that an increase in permeability and hydraulic conductivity leads to an increase in drug concentration in the tumor. Finally, Methotrexate drug has the most effect in the treatment of tumor. Overall, the computational model for anti-cancer delivery provides a powerful tool for understanding and optimizing drug delivery strategies for the treatment of cancer.

# 1. Introduction

The development of computational models for the correct prediction of drug delivery and treatment, its improvement and optimization, and the introduction of new treatment methods can help personalize treatment and increase patients' life expectancy and quality of life [4]. Computational modeling of drug delivery to specific tissues has a relatively long history [35-39] as providing valuable results that contribute to our understanding of disease progression and treatment. The power of mathematical modeling lies in its ability to reveal previously unknown physical principles that may have been overlooked in the qualitative approach to biology. This has led to many successes in various fields related to various models of interaction between cells and their microenvironments [34].

The study of drug delivery to different tissues includes studies in two areas related to fluid flow in the vascular and capillary networks and drug delivery in the interstitial space. Over the past three decades, Jane et al. have taken basic steps using fluid mechanics principles in drug delivery studies [5-10]. In their research, the interstitial fluid flow has been investigated, and the interstitial fluid pressure has been considered an influential factor. In their view, body tissue is a porous material, and its lower levels, like a capillary network, are solved by simple assumptions. This method is a macroscopic view of modeling fluid flow in tumor tissue. Soltani et al. [11] developed a mathematical interstitial flow model and applied their model to various geometries of tumor tissue to study the effect of tumor shape and size [12-13]. In these studies, the capillary network effect was replaced by a constant pressure uniformly distributed within the tissue. Conventional drugs in chemotherapy to tumor cells involve injection and transfer through the circulatory system, seepage from the vessels around the tissue, and interstitial transfer to reach the target cells, bind to them, and enter the cells for treatment. The efficiency of the drug delivery process depends on the spatio-temporal distribution of the drug in the target tissue, which depends on the characteristics of the tissue environment and the drug. Therefore, several parameters play a role in this process, some related to the drug and some to the tissue. In the study by Moradi Kashkoli et al. [1], The effect of the drug release rate parameter ($K_{el}$), drug binding rate parameter to cell ($K_{ON}$), a dose of an injected drug, investigated. Moradi Kashkoli et al. [1] and Soltani et al. [12], also, analyzed the effect of interstitial fluid pressure and the shape and size of tissue on drug delivery. (In the field of drug delivery, soluble materials can be transferred through porous materials using mechanisms of material transfer and capillaries [18]. Jane et al. proposed a model of fluid transfer and antibody distribution in a tumour's tissue with a simplified geometry using the finite element method [5-6]. From the late 1990s until today, several researchers have presented and reviewed simulation studies of drug distribution in tumor and healthy tissues to solve various problems related to drug delivery [26-30]. Capillary geometry has recently been used to increase the accuracy of fluid flow calculations and drug distribution. For example, Sefidgar et al. [19-21] presented a model that incorporates various physiological factors of drug delivery to the tumor by considering the dynamic capillary network. By modeling the geometry of the capillary network, Ji Woo et al. [14] developed a mathematical model of the capillary flow system and its intermediate flow coupling through the permeability of the vessels.

By modeling the geometry of the capillary network, Ji Woo et al. [14] developed a mathematical model of the capillary flow system and its intermediate flow coupling through the permeability of the vessels. Shojaei and Niroo Oskooi [22-23], Soltani et al. [24], and Asgari et al. [25] examined the distribution of chemotherapy and tracer in the tumor, respectively, by considering the artificial capillary network in the tumor. A review of the literature on this subject showed that the effect of various parameters and modeling conditions on drug delivery had been discussed. However, so far, no research has examined the two parameters of permeability and hydraulic conductivity in drug delivery.

In the geophysical studies on fluid flow in porous media, hydraulic conductivity is a well-described characteristic that regulates fluid invasion, flow rates, and pore fluid pressure distribution. A solid tumor can be regarded as a porous medium, and hydraulic conductivity plays a crucial role in the majority of solid tumors increase in interstitial pressure [40]. A further important parameter is vascular permeability, as vascular permeability not only affects tumor biology, but also plays a crucial role in delivering macromolecular anticancer drugs to tumors [41]. By developing models to measure and manipulate hydraulic conductivity and permebility of vessels, researchers may be able to improve drug delivery to the tumor microenvironment and improve the success of chemotherapy regimens. In this study, for first time, using multiscale mathematical modeling based on the real image by considering different states for two parameters, permeability and hydraulic conductivity and also assumes different real conditions in the input and output parent vessels to target tissue, for three drugs Doxorubicin, Cisplatin and Methotrexate, has been studied.

## 2. Material and Methods

To begin modelling drug delivery to tissues, the drug is injected into the circulatory system and subsequently finds its way into the interstitial area through the vessels, finally moving into the ECM (Extracellular Matrix). The drug is then driven into the target area via the binding affinity between drug ligands and cell receptors. Next, cellular uptake and drug internalization continue this process. In plain terms and through the perspective of a continuous model, the drug delivery process entails transporting the drug to the TTE (Targeted Tissue Microenvironment), moving in the ECM, binding to the cells, and ultimately internalizing within the cells [12]. Figure 1 includes the schematic of a normal drug delivery system, along with elaborations on the interstitial space, the capillary network, the lymphatic system, drug transport convection and diffusion mechanisms, and the drug delivery to the intravascular, interstitial, and intracellular spaces. Based on the adopted approach, the problem addressed in this research is multiscale as it involves the scales of nanometers, micrometers and centimeters for cellular uptake and binding, blood flow through the vascular network, and fluid flow and solute transport in the interstitium, respectively [21]. The subsequent sections entail thorough investigations of the governing equations, computational field, modeling parameters, boundary conditions, geometry, solution strategy, and result validation.

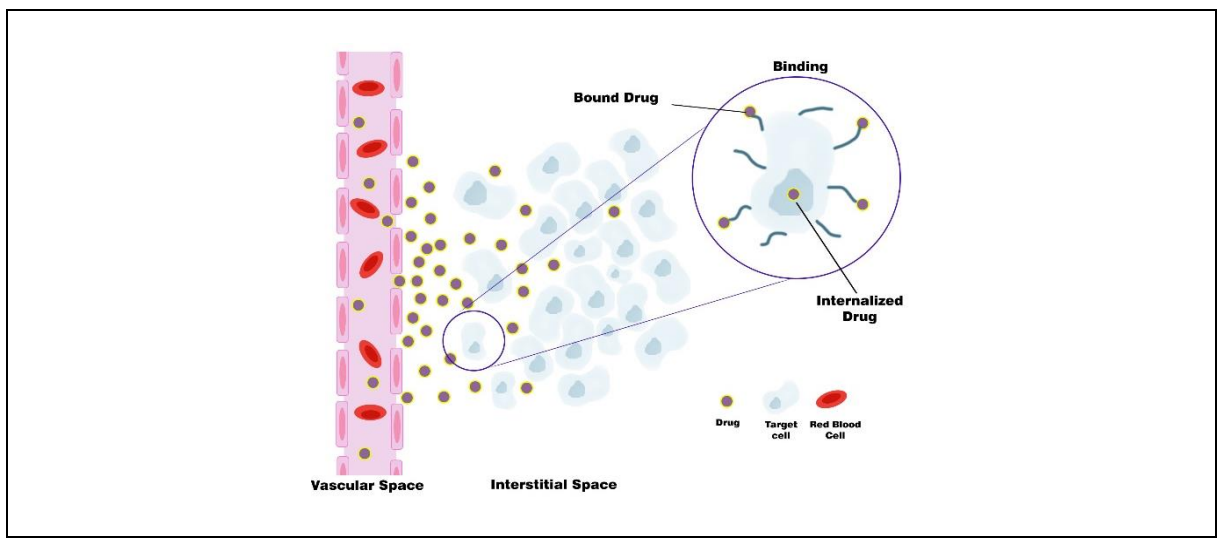

**Figure 1.** A basic system of drug delivery. The schematic representation includes the interstitial space, the capillary network, the lymphatic system, diffusion and convection transport mechanisms, and drug transport in intracellular, interstitial, and intravascular spaces.[1]

## 2.1. Governing Equations

This study thoroughly explains the governing equations related to drug transport and distribution and interstitial and intravascular fluid flows.

### 2.1.1. Intravascular Velocity

Poiseuille's law (also applicable in low Reynolds numbers) is used to discover the intravascular velocity inside the capillary network. Notably, the Poiseuille law was modified in this context according to the blood physics operating in the vasculature, which is heavily influenced by blood viscosity and hematocrit [21, 33]. Additionally, the evident properties of the blood were acquired from Sa'adi pour *et al.* [2]:

As shown below, the real physiological conditions expressed in [32] were taken into account to select inlet and outlet pressure values of the parent vessels:

$P_{Inlet,1} = P_{Inlet,2} = P_{Inlet,3} = P_{Inlet,4} = 25$ mmHg     $P_{Outlet,1} = P_{Outlet,2} = P_{Outlet,3} = 10$ mmHg

A flux was presented based on Starling's law to obtain the fluid transport rate from vessels to tissues which is demonstrated below [12]:

$$Q_t = \pi D L L_p \left( P_b - P_i - \sigma_s (\pi_b - \pi_i) \right) \quad (1)$$

in which $Q_t$ is the trans-vascular blood flow rate, $L$ is the vessel wall length, $L_p$ is the hydraulic conductivity of the vessel wall, $D$ is the capillary diameter, $P_i$ is IFP, and $P_b$ is the intravascular blood pressure. In addition, $\sigma_s$, $\pi_b$, and $\pi_i$ and are the mean of the osmotic reflection coefficient for plasma proteins and the osmotic pressures of the capillaries and the interstitial fluid, respectively.

### 2.1.2. Interstitial Fluid Flow

The Darcy equation [4] was used to simplify the momentum model governing fluid flow in biological tissues. Based on this equation, the IFV (Interstitial Fluid Velocity) relies on the gradient of pressure; consequently, [31]:

$$\nabla P_i = -\left(\frac{\mu}{k}\right) v_i \quad (2)$$
$$v_i = -\kappa \nabla P_i \quad (3)$$

where $\kappa$ is the tissue hydraulic conductivity.

Conversely, given the existence of the source (referred to here as 'the capillary network') and the sink terms (referred to, here, as 'the system of lymphatic drainage), the continuity equation is corrected in the form of the following [16]:

$$\nabla \cdot v_i = \phi_B - \phi_L \quad (4)$$

where $\phi_B$ determines the fluid flow rate from the vascular network to interstitial space, and $\phi_L$ indicates the fluid flow rate from tissue to the lymphatic drainage system.

Starling's law was employed to calculate the rate of trans-vascular flow, specifying hydrostatic and osmotic pressures while flowing through the capillary membrane. As a result, [16]:

$$\phi_B = L_P \left(\frac{S}{V}\right) \left( P_b - P_i - \sigma_s (\pi_B - \pi_i) \right) \quad (5)$$

where $S/V$ represents the surface area per micro-vascular network unit volume, the net filtration rate is obtained from the value of this equation.

The equation below can be used to acquire the flow rate of the lymph, which is regarded as uniform across a typical tissue [16]:

$$\phi_L = L_{PL} \left(\frac{S}{V}\right)_L (P_i - P_L) \quad (6)$$

in which $L_{PL}(\frac{S}{V})_L$ is the lymphatic filtration coefficient, and $P_L$ is the hydrostatic pressure of the lymph.

Integrating Equations (4) and (3), while maintaining $\kappa$ as constant, leads to:
$$-\kappa \nabla^2 P_i = \phi_B - \phi_L \tag{7}$$
Previous research has accurately laid out the method for IFP calculation [12, 21].

### 2.1.3. Drug Transport in Porous Medium

Equations related to drug delivery in the interstitial space entail the mechanisms for convection and diffusion. Notably, the lymphatic system and the capillary network are regarded as sink and source terms. Taking Fick's law into account, the CDR (Convection-Diffusion-Reaction) system of equation, respectively ns was employed for drug transport as shown below [8, 32]:

Free drug:
$$\frac{\partial C_F}{\partial t} = -\nabla \cdot (v_i C_F) + \nabla \cdot (D_{eff} \nabla C_F) - \frac{1}{\varphi} K_{ON} C_{rec} C_F + K_{OFF} C_B + (\Phi_B - \Phi_L) \tag{8}$$
Bound drug:
$$\frac{\partial C_B}{\partial t} = \frac{1}{\varphi} K_{ON} C_{rec} C_F - K_{OFF} C_B - K_{INT} C_B \tag{9}$$
Drug internalized into the cell:
$$\frac{\partial C_{INT}}{\partial t} = K_{INT} C_B \tag{10}$$

in which, as the signs suggest, $C_F$, $C_B$, and $C_{INT}$ signify the concentrations of free, bound, and internalized drugs, respectively. $C_{rec}$ represents the drug's concentration at the cell surface of the receptor. In addition, $v_i$ is the IFV, calculated via the Darcy equation, $D_{eff}$ is the coefficient of diffusion in tissue. $K_{ON}$, $K_{OFF}$, and $K_{INT}$ respectively represent the constants regarding the rate of drug binding to cell receptors, unbinding, and internalization into the cell through the receptors. $\varphi$ signifies the volume fraction of the targeted tissue accessible to the drug. Finally, $\Phi_B$ and $\Phi_L$ are the drug exchange rates from capillaries to the interstitium and from the interstitium to the lymphatic system, respectively. Furthermore, to calculate and measure the trans-vascular rate from the capillary wall, the Patlak model [15] was applied. Ultimately, the transport rate of solute from capillaries to interstitium was found as represented below [8]:
$$\Phi_B = \phi_B (1 - \sigma_f) C_P + \frac{PS}{V}(C_P - C_F)\frac{Pe}{e^{Pe} - 1} \tag{11}$$
$$Pe = \frac{\phi_B (1 - \sigma_f) V}{PS} \tag{12}$$

where $Pe$ is a Péclet number, representing the convection rate's ratio to diffusion from the capillary wall, $\sigma_f$ is the filtration reflection coefficient, $P$ is the permeability of capillaries, and $C_P$ is the concentration of drug in plasma.

Presumed as an unvarying distribution, exclusively in a typical tissue, the rate of solute transport through the lymphatic system was considered as the following [12]:
$$\Phi_L = \phi_L C_F \tag{13}$$

## 2.2. Geometric Model and Computational Domain

In the present research, a two-dimensional model built upon a portion of a real tissue image with a capillary network was used as input, according to Moradi Kashkooli *et al.* [1]. A 2D computational field was taken into account, following image extraction, along with the presence of an elliptical domain in the middle of the field and the parent vessels (Fig. 2). The simulation was prepared, and then various extents of the primary parameters, such as the hydraulic conductivity and the permeability of vessel wall was examined for Doxorubicin, Cisplatin and Methotrexate, which are known as three conventional drugs; the next section includes the obtained results.

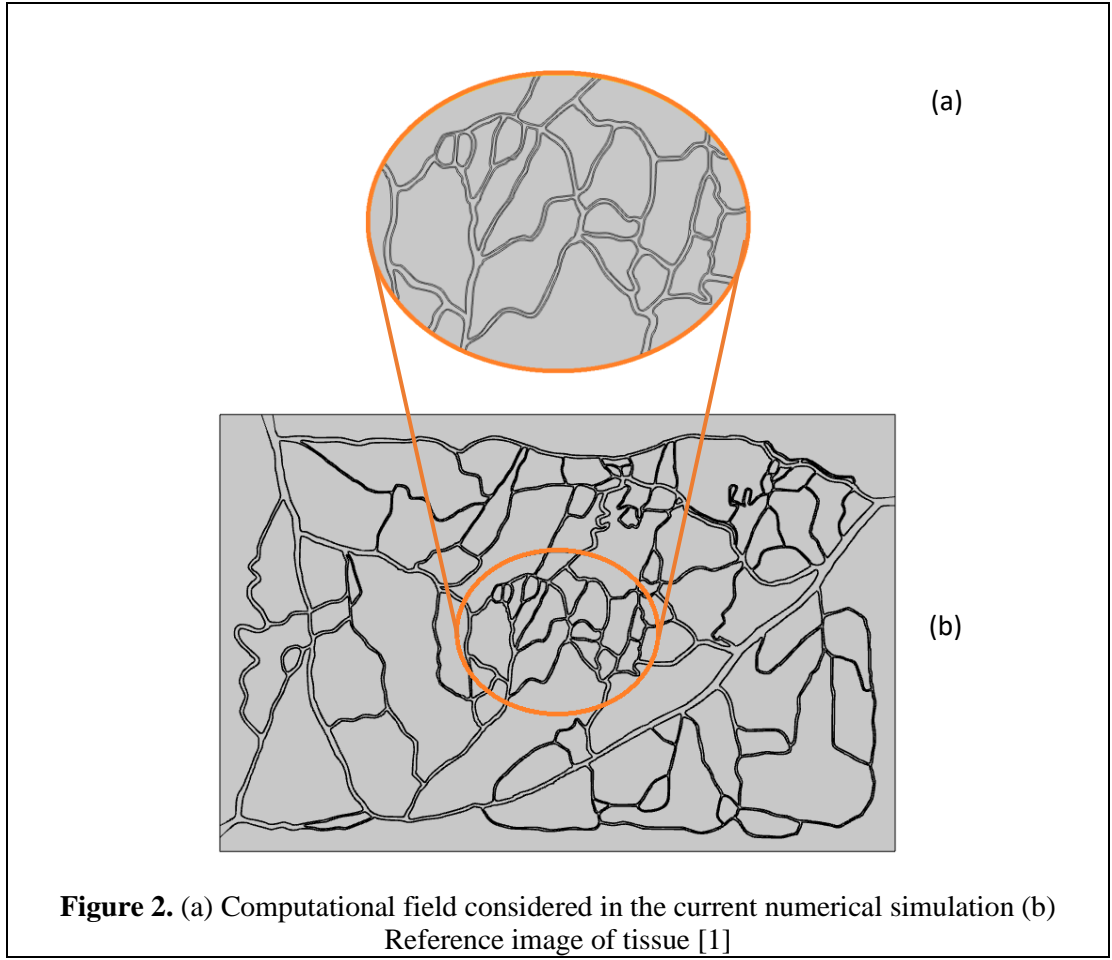

**Figure 2.** (a) Computational field considered in the current numerical simulation (b) Reference image of tissue [1]

### 2.3. Boundary Conditions and Grid Independency

Two boundaries were assumed as identical for normal tissue, according to Moradi Kashkooli *et al.* [1]. Table 1 presents the governing boundary conditions used on the interstitial fluid flow and the solute transport equation.

**Table 1.** The governing boundary conditions in this study.

| Area | Boundary conditions | |
|---|---|---|
| | Fluid flow | Solute transport |
| Center of the Geometry | $\nabla P_i = 0$ | $D_{eff}\nabla C + v_i C = 0$ |
| External boundary | $P_i = $ Constant | $-n \cdot \nabla C = 0$ |

The chemotherapeutic drugs' mass injection was modeled in the present research via an approach by which the concentration was reduced at the inlet of the vascular network; as a result [3]:

$$C_P = \exp(-t/K_d) \tag{19}$$

where $K_d$ represents a time constant demonstrating the drug's half-life in plasma.

**Mesh Independency**

Table(2). Different states of grid independence.

| Mesh | domain elements | boundary elements | Concentration ($mol/m^3$) | Geometry |
|---|---|---|---|---|
| Fine | 844689 | 73646 | 0.26569 | 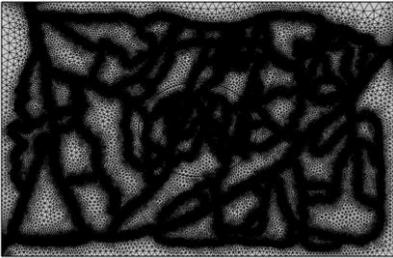 |
| Extra Fine | 1709664 | 106472 | 0.26689 | 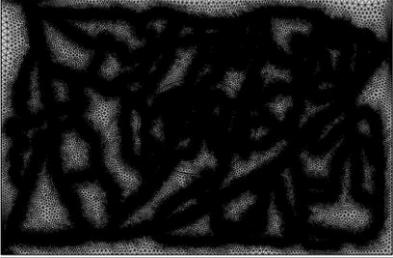 |
| Extremely Fine | 2391327 | 112075 | 0.26697 | 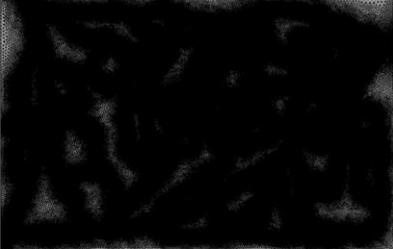 |

To evaluate the network's independence for the present study, three different meshes, including fine, finer and extremely fine grids were implemented on the domain, each of which can be seen in Table (2). They created networks are of triangular type, created using Comsol Multi-Physics software, and the final result of the solution can be seen. As shown in the Table, the concentration of internalized drugs in the cells, $C_{INT}$, which is the output parameter of the modeling, in the finer and extremely fine state, respectively, has 0.0012 and 0.00128 changes. Due to the hardware configurations costs as well as the time required to solve, the fine state is considered as the reference in the whole research

**2.4. Numerical Approach and Values of Parameters**

To acquire blood velocity and pressure in the capillary network, a Poiseuille equation was employed [17, 21]. The Interestitial and intra-vascular pressure were integrated via the Starling law, demonstrating an extravasation rate from the capillary membrane. Next, the Darcy equation was used to calculate the interstitial fluid flow, and the achieved values of Interstitial Fluid Velocity (IFV) and Interstitial Fluid Pressure (IFP) were applied to solve the equations related to drug transport. Solving the equations of the transient solute transport in numerical terms took 60 hours with a time step of 0.01. Notably, $1 \times 10^{-6}$ was the numerical precision of the responses. The commercial finite element software, COMSOL Multiphysics (Ver. 6.0), was used to carry out the entire simulations. The computer's specifications for the simulations involved an Intel processor (2.8 GHz - 7[th] generation) and 16 GB of memory.

In addition to DOX, Cisplatin and Methotrexate were considered in this study since it is one of the most typical chemotherapeutic drugs. Notably, the numerical approach provided here can also be used for any drug. Table (3) lists the drug characteristics in the target tissues and Table (4) presents the necessary values for solving the interstitial fluid flow.

**Table (3).** characteristics of the drug in the target tissue

| Parameter | Unit | Dox | Cis | Meth | Reference |
|---|---|---|---|---|---|
| D_normal | m²/s | 3.4×10⁻¹⁰ | 2.46×10⁻¹⁰ | 1.81×10⁻¹⁰ | [42] |
| D_tumor | m²/s | 1.58×10⁻¹⁰ | 5.27×10⁻¹⁰ | 3.8×10⁻¹⁰ | [8] |
| $P$ | m/s | 3.75×10⁻⁷ | 3.75×10⁻⁷ | 3.75×10⁻⁷ | [8] |
| $\sigma_f$ | - | 0.35 | 0.35 | 0.35 | [12] |
| $K_{ON}$ | 1/(M·s) | 1.50×10⁴ | 1.50×10⁴ | 1.50×10⁴ | [12] |
| $K_{OFF}$ | 1/s | 8×10⁻³ | 8×10⁻³ | 8×10⁻³ | [12] |
| $K_{INT}$ | 1/s | 5×10⁻⁵ | 5×10⁻⁵ | 5×10⁻⁵ | [12] |
| $\varphi$ | - | 0.4 | 0.4 | 0.4 | [25] |
| $C_{rec}$ | M | 1×10⁻⁵ | 1×10⁻⁵ | 1×10⁻⁵ | [25] |
| $K_d$ | min | 6 | 6 | 6 | [12] |

**Table (4).** Parameters used for solving the flow of the interstitial fluid

| Parameter | Unit | Value | Reference |
|---|---|---|---|
| $\pi_b$ | mmHg | 20 | [4] |
| $\pi_i$ | mmHg | 10 | [4] |
| $\sigma_s$ | - | 0.91 | [4] |
| $L_p$ | cm/((mmHg)*s) | 0.36×10⁻⁷ | [4] |
| $\kappa$ | cm²/(mmHg*s) | 8.53×10⁻⁹ | [4] |
| $L_{pL}S_L/V$ | 1/(mmHg*s) | 1.33×10⁻⁵ | [27] |
| $P_L$ | Pa | 0.0 | [27] |

**Validation**

To validate the current study, we compared the results of the baseline state with Farshad Moradi Kashkooli et al. work [1]. In mentioned work, the parameter $C_{INT}$ in tumor tissue is considered as one of the final outputs of the study. In figure (3), the trend of drug distribution over 60 hours after injection is demonstrated, which has a flat behavior of around 0.26 $Mol/m^3$. Also, in the present study, $C_{INT}$ is exactly 0.26, which is similar to the mentioned study. Due to differences in stady and various inputs and outputs, the difference between two amounts of $C_{INT}$ is occurred.

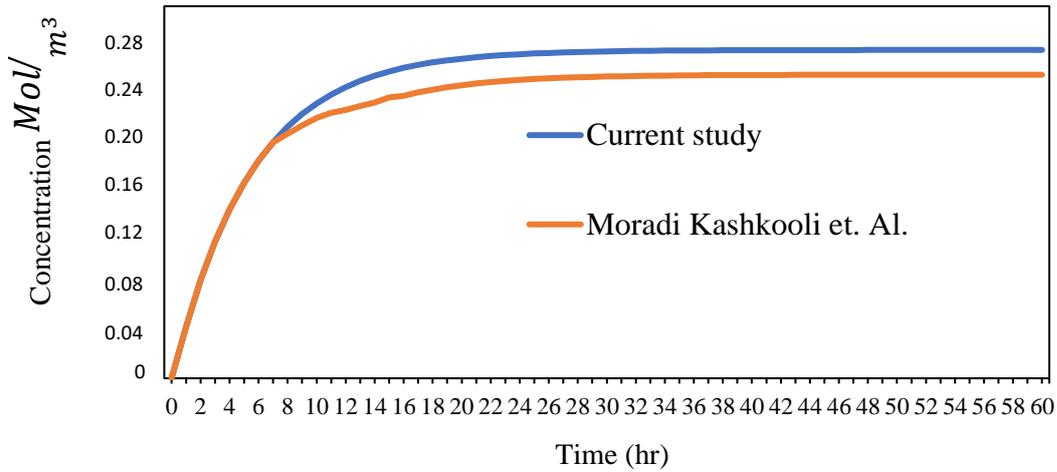

**Figure (3)**. Validation of presented result with Moradi Kashkooli et al. [1]

As shown, the results have similar approaches, and they matched each other during the time till hour 60. So, it can be understood that the mathematical modeling and geometry in the present study is validated by the mentioned work and can be implemented for other drugs.

**Drug information**

To do this project, the distribution of three different drugs, Doxorubicin, Cisplatin and Methotrexate, in the domain is investigated. The information on each can be seen in Table (5).

**Table (5).** Information on three different drugs used in the current study.

| Drugs | Molecular Weight ($g/Mol$) | Reference |
|---|---|---|
| Doxorubicin | 554 | (Melik-Nubarov and Kozlov, 1998) |
| Cisplatin | 300 | (Saltzman and Radomsky, 1991) |
| Methotrexate | 454 | (Saltzman and Radomsky, 1991) |

The present study investigated the effect of two vital parameters (hydraulic conductivity and permeability) on IFV, IFP, and drug concentration. The range of investigated parameters is as follows:
The permeability in normal tissue: 3.75e-5 to 1e-2
The permeability in tumor: 3e-4 to 1e-1
Hydraulic conductivity in normal tissue: 2.7e-12 to 1e-5
Hydraulic conductivity in normal tissue: 2.1e-11-1e-5

**Results**

Figure 4 depicts the spatial variations of Dox, Cisplatin, and Methotrexate concentrations in the intracellular space throughout a 60-hour period. The Cisplatin medication has a larger concentration than the other two pharmaceuticals during this time, according to the findings. Cisplatin has a lower molecular weight, which enables it to be more effective in areas with denser blood vessels than the other two drugs. The findings show that the tumor has greater levels of all three medicines than the surrounding normal tissue. Aside from this, it is also notable that Dox caused the least damage to the normal tissue, despite the fact that it triggers cell death in the tumor, though it has an effect on normal tissue as well.

Therefore, the results indicate that the increase in molecular weight has a great effect on IFP (figure 4). Cisplatin, with a molecular weight of 300 grams/mol, has a pressure of the lowest pressure. The highest pressure is associated with Doxorubicin, which has a molecular weight of 554 grams/mol.

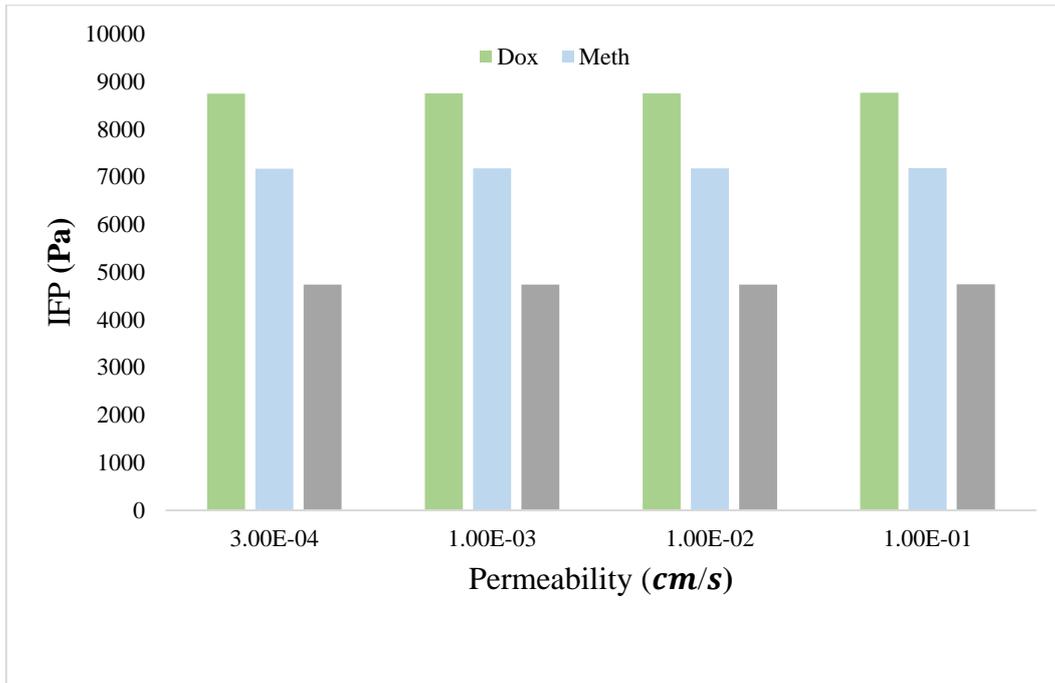

**Figure 4**. Effect of different values of permeablity on IFP in tumor tissue

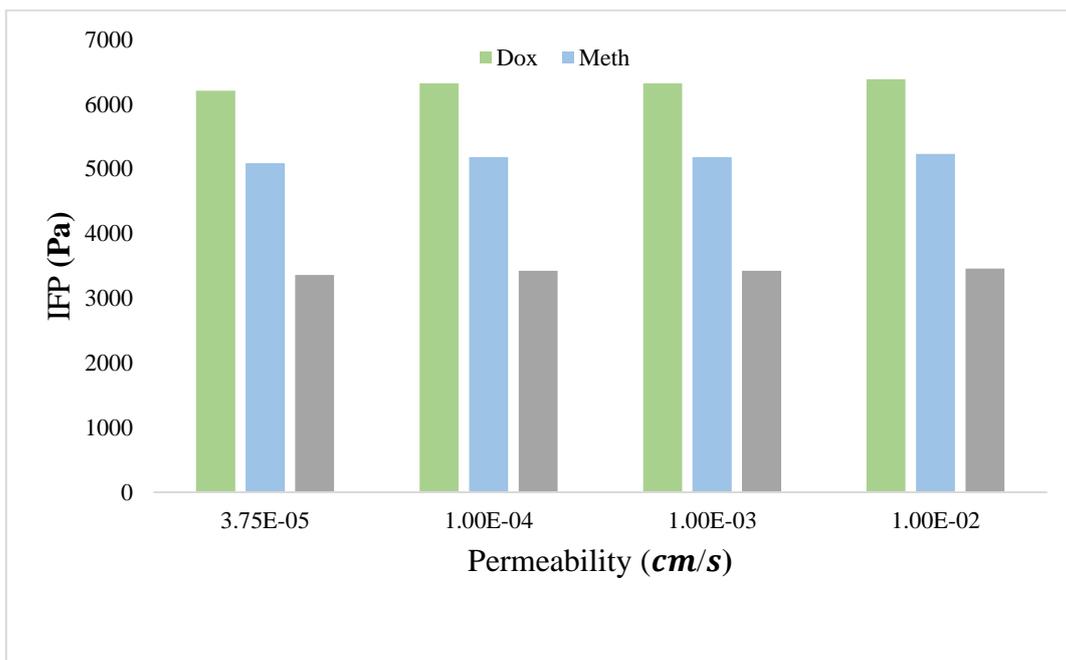

**Figure 5**. Effect of different values of permeablity on IFP in normal tissue

**Permeability**

Figure 4 and 5 shows the effect of increasing permeability on interstitial fluid pressure for three drugs: Doxorubicin, Cisplatin and Methotrexate. As shown in these figures, increasing permeability increases the interstitial fluid pressure, however, these changes are not significant. The same behavioral pattern was observed for all three drugs. The results of this study confirm previous findings that the tumor pressure is higher than normal tissue pressure. In the case of Dox, for example, the pressure inside the tumor is around 9000 Pa, while the pressure inside the normal tissue is around 6300 Pa.

The effect of increased permeability on the concentration of all three drugs in tumor and normal tissue has been studied. The figure 6 illustrates that an increase in vascular permeability from 10e-4 to 10e-1 results in twice as much increase in the concentration of a drug in the tumor.

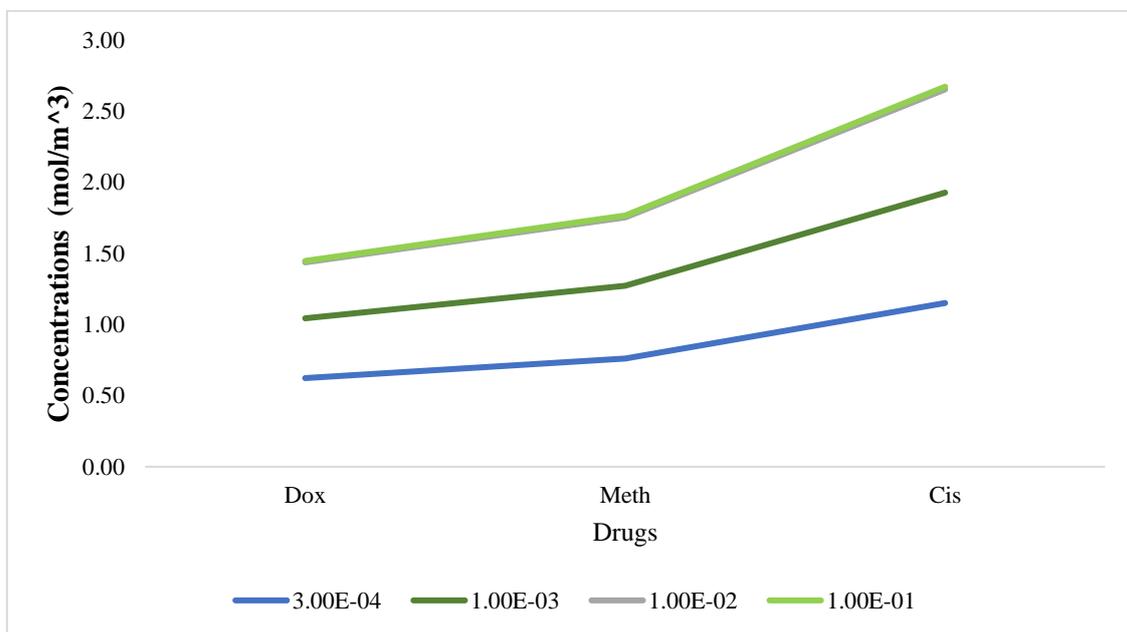

**Figure 6**. Concentration of each drug for different values of Permeability in tumor tissue

**Hydraulic conductivity**

The effects of changes in hydraulic conductivity on the distribution of drugs over a range of 2E-11 to 1E-5 m.Pa/s have been investigated. Furthermore, to the effect on drug concentration and distribution, hydraulic conductivity can also have an influence on IFV and IFP within the tumor microenvironment. Based on the results (figure 7), hydraulic conductivity impacts interstitial fluid pressure significantly. It is evident from the results for all three drugs that an increase in hydraulic conductivity results in pressure decrease. For example, increasing hydraulic conductivity from 2.1E-11 to 1E-2 caused the pressure to decrease by 2464Pa (Doxorubicin), 2019.23 Pa (Cisplatin), and 1334.3 Pa (Methotrexate) in tumor. Drug distribution in tumors and normal tissues can be affected by these changes.

There is evidence that changes in hydraulic conductivity in normal tissue and tumor tissue have a significant effect on concentration (figure 8). For the drug Dox, Cisplatin and Methotrexate, it has increased the concentration from 109.26 to 0.62, 1.15 to 201.77, 0.76 to 133.33, respectively.

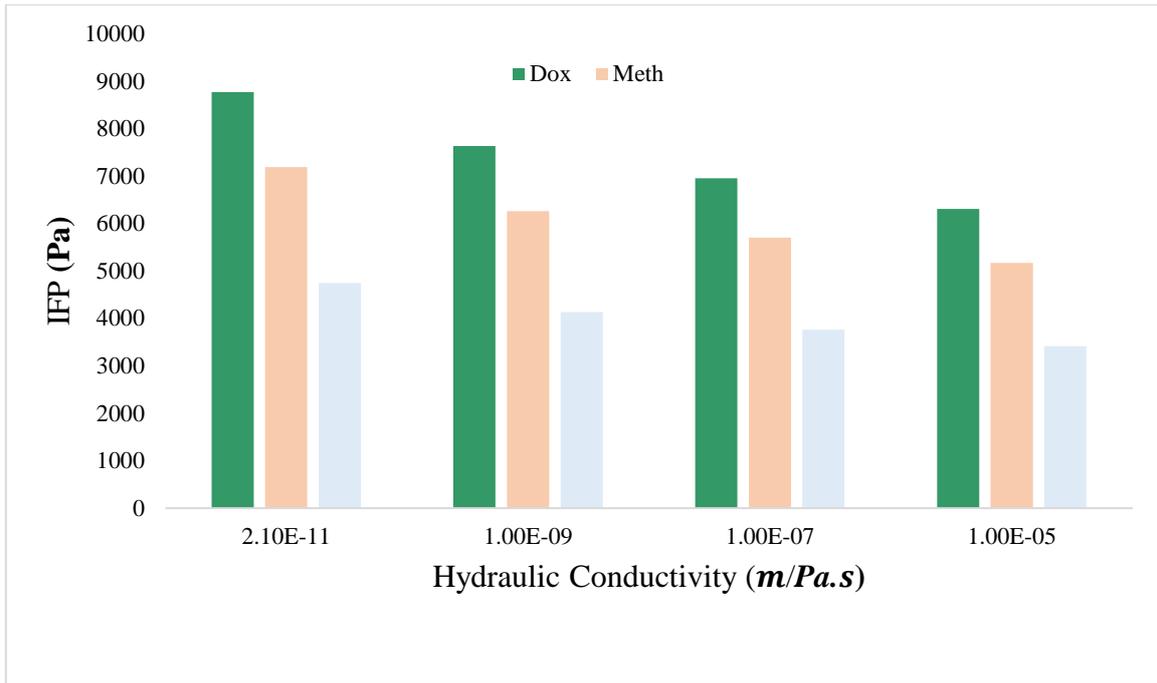

**Figure 7**, Effect of different values of Hydraulic Conductivity on IFP in Tumor tissue

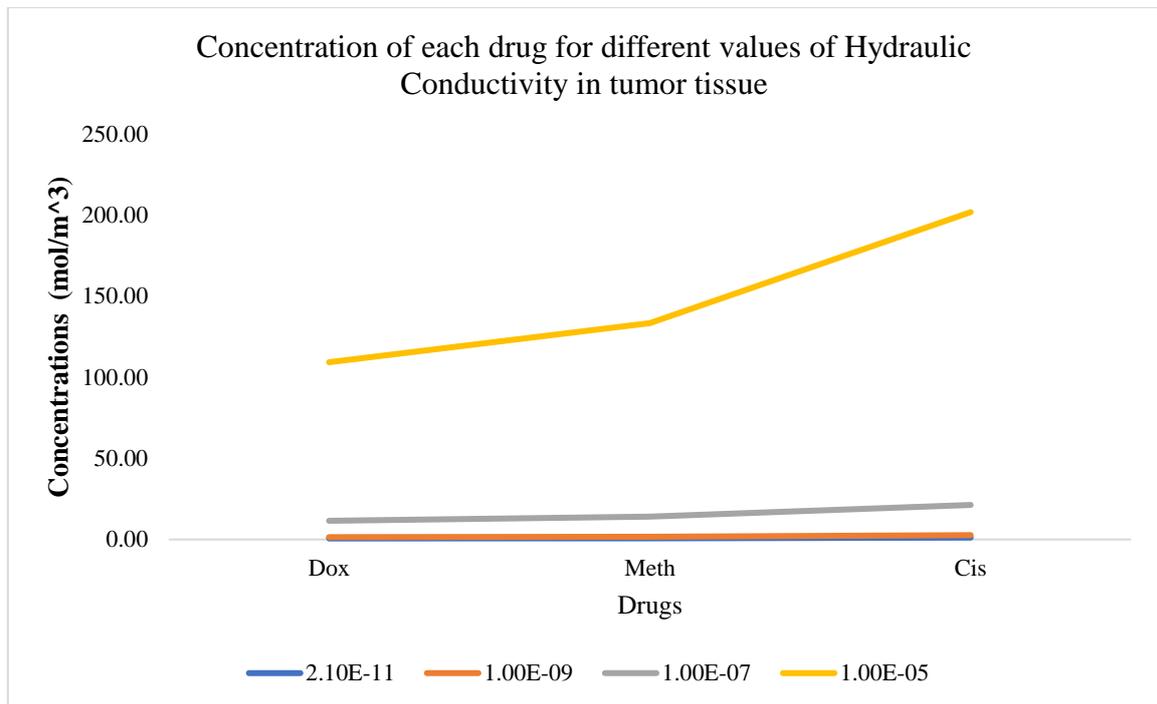

**Figure 8.** Concentration of each drug for different values of Hydraulic Conductivity in tumor tissue

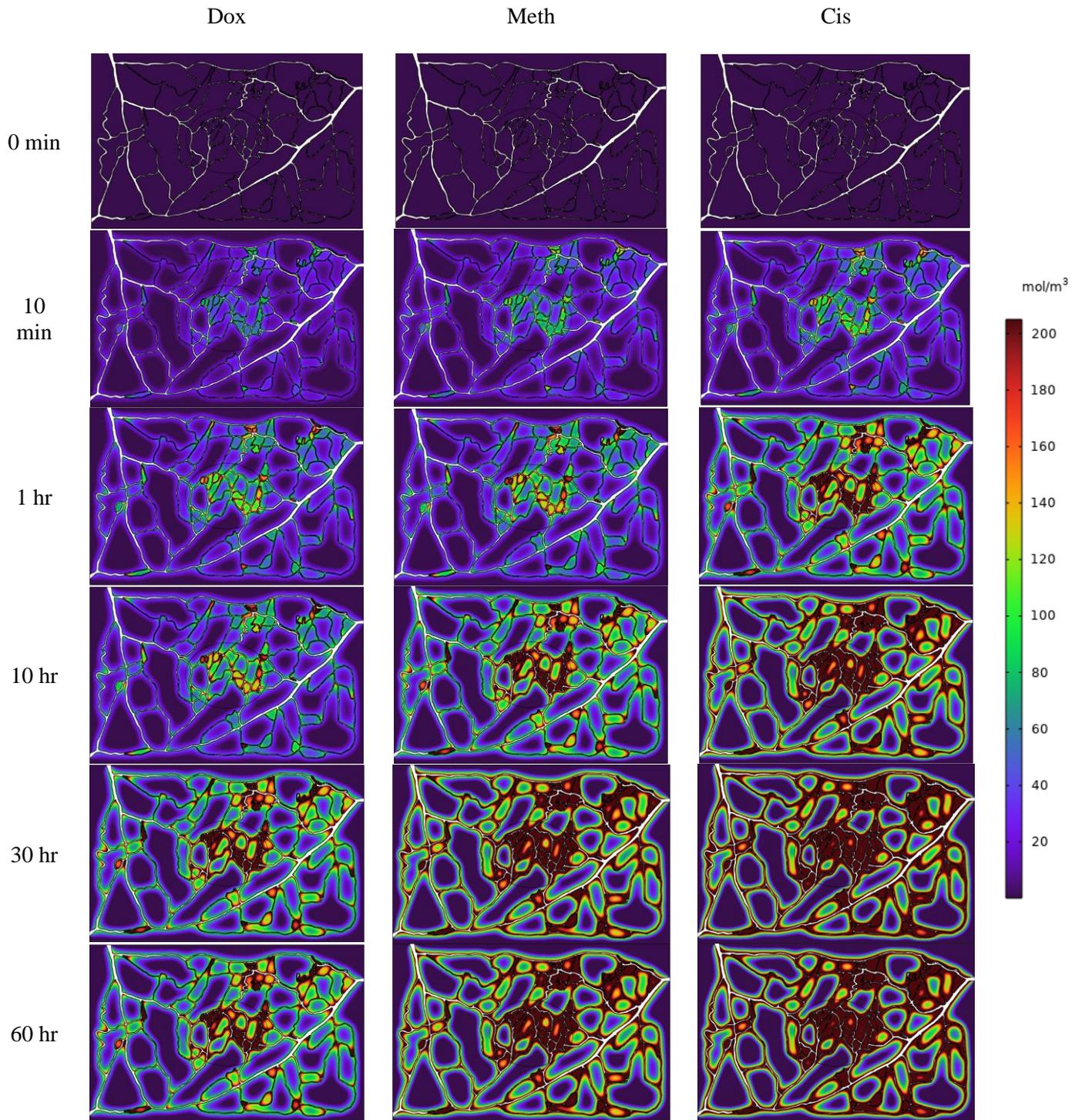
**Figure 9.** Spatial concentration of Dox, Cisplatin, and Methotrexate during 0 to 60 hours

**Discussion**

As part of this investigation, for the initial time, a model has been constructed to analyze the impact that parameters Hydraulic conductivity and permeability have on concentration by making use of a real image. A comprehensive approach was constructed to study the delivery and accumulation of Dox, Cisplatin and Methotrexate in solid tumors. Mathematical models can be used to investigate the effect of vessel permeability, hydraulic conductivity, and types of anticancer agents on IFP, concentrations, in tumors. Figure 9 compares the drug concentrations of the three medicines in the intracellular space. Compared to three drugs, Cisplatin has a greater effect when it comes to treating tumors (1.15 $mol/m^3$). The molecular weight of drugs and drug delivery systems can have a significant effect on their efficacy and pharmacokinetics. In general, smaller molecules can diffuse more easily through tissue and cell membranes, leading to faster uptake and distribution within the body [43]. In contrast, larger molecules may have more limited penetration and uptake, resulting in slower clearance and longer circulation times.The results of the present model show that Cisplatin distributes more quickly and receives a higher concentration in the tumor as Cisplatin has the lowest molecular weight.

Hydraulic conductivity is an indicator of the ease with which fluids can flow through a tissue. In the context of tumor biology, hydraulic conductivity measures the permeability of the extracellular matrix and capillaries within the tumor, which can influence the distribution of chemotherapy drugs. Hydraulic conductivity can impact IFP by affecting the rate of fluid flow into and out of the tumor microenvironment. Tumors with low hydraulic conductivity may have higher IFP due to reduced fluid flow and increased resistance to fluid movement. In contrast, tumors with high hydraulic conductivity may have lower IFP due to enhanced fluid flow and reduced resistance. The results of the present study show an increase in hydraulic conductivity leads to a decrease in IFP which is compatible with previous studies [17], [44], [45], [46]. Lambride et al [46] used in siloco model to study distribution anti cancer drug. They showed, when the tumor interstitial space hydraulic conductivity declines, tumor velocity drops and tumor center fluid pressure rises. The permeability of vessels is another parameter that was studied in the study. The high permeability of vascular networks can cause fluid and proteins to leak into the tumor interstitium and elevate IFP. The results indicate that the increase in permeability caused an increase of approximately 100 Pa for all three drugs. Over all, the results of this study show that IFP is more affected by hydraulic conductivity than permeability.

Cancer treatment success is related to the balance between permeability and IFP, since it can affect drug distribution and efficacy in the tumor. There are three drugs whose concentration can be increased by an increase in permeability. Dox, for instance, has a high molecular weight, which allows it to penetrate into the interstitial space more easily.

A major limitation of the study was the number of drugs, different tissues and the various values of study conditions. In this work,in addition to Dox, Cisplatin and Methotrexate, other drugs can be investigated and their trends can be discussed. Furthermore, these modeling, can be implemented on other tissues, IFV, IFP and drug concentration can be examined. Also, in this study, four states of hydraulic conductivity and four ones of permeability has been checked. The modeling can be investigated in other states to show the results.

In further research, the therapy of all three medications at the same time will be examined jointly. Furthermore, various treatment protocols may be investigated with the help of this model.

## Conflict of Interest

The authors of present study declare they do not have any conflict of interest.

## Authors' contribution

**Masod Sadipour:** modeling, reviewing, writing, visualization.

**M.Masoud Momeni:** conceptualization, methodology, modeling, discussion, edition, software.

**Madjid Soltani:** supervision, project administration, review, and edition.

## Authorship contribution